\documentclass[11pt]{article}

\usepackage{mathptmx}

\usepackage{amsmath,amssymb} 

\newtheorem{remark}{Remark}[section]

\newtheorem{lemma}{Lemma}[section]

\newtheorem{theorem}{Theorem}[section]
\newtheorem{proposition}{Proposition}[section]

\newtheorem{corollary}{Corollary}[section]

\newcommand{\proof}{\textsc{proof}\quad}
\newcommand{\qed}{\hfill \textsc{qed}}

\numberwithin{equation}{section}

\begin{document}
\title{Quantum energy-mass  spectra of relativistic Yang-Mills fields in a functional paradigm}
 
\author{Alexander  Dynin\\
\textit{\small Department of Mathematics, Ohio State University}\\
\textit{\small Columbus, OH 43210, USA}, \texttt{\small dynin@math.ohio-state.edu}}

\maketitle

\begin{abstract}
A  non-perturbative and mathematically rigorous  quantum Yang-Mills theory on 4-dimensional Minkowski spacetime is set up in  the functional framework of a complex nuclear  Kree-Gelfand triple. It involves a symbolic calculus of operators with variational derivatives and  a new kind of infinite-dimensional ellipticity.
In the temporal gauge and Schwinger first order formalism,  Yang-Mills equations  become a semilinear hyperbolic system for which the general Cauchy problem  is reduced to  initial data with compact supports.  For a simple compact Yang-Mills gauge group and  the anti-normal quantization of  Yang-Mills energy-mass functional of  initial data in a box,  the quantum energy-mass spectrum  is a  sequence of non-negative eigenvalues converging  to infinity. In particular, it has a positive mass gap.  Furthermore, the energy-mass spectrum is   self-similar (including the mass gap) in  the  inverse proportion to  an infrared cutoff  of the classical energy scale.  

 \medskip
\emph{Key words}: Second quantization; Kree-Gelfand nuclear triples; operators with variational derivatives; 
infinite-dimensional symbolic calculus;  infinite-dimensional ellipticity; Yang-Mills Millennium problem;  asymptotic freedom.

 \medskip
2010  AMS Subject Clasification: 46E50,  81T13, 47G30, 35P15,

\end{abstract}

\begin{flushright}
\textit{In memoriam I. M. Gelfand on  his 100th anniversary.}
\end{flushright}

\section{Introduction}

As proven in this paper, the quantum energy-mass spectra of relativistic Yang-Mills fields with infrared  cutoffs are  infinite and discrete. Moreover these  spectra are self-similar in an inverse proportion to  the classical energy scale,
so that the mass gap converges to zero the cutoff goes to infinity/

Presumably this provides   a solution for \emph{both} parts of the 7th Millennium  problem of Clay Mathematics Institute: 
\begin{quotation}
\textsl{Prove that for any compact simple global gauge group, a nontrivial  quantum Yang-Mills theory exists  on Minkowski time-space and has a positive mass gap} (cp. \textsc{jaffe-witten}\cite{A}).
\end{quotation}
(This formulation is from \textsc{witten}\cite[p. 24]{Witten-02}.) \footnote{The official formulation of the Millennium Yang-Mills problem \cite{A} is looking for   a quantum Yang-Mills theory with  axiomatic properties at least as strong as  axioms of Wightman  relativistic  quantum field theory. However   even modified  Wightman  axioms  (see, e.g., \textsc{bogoliubov} et al \cite[chapter 10]{Bogoliubov} are in a serious conflict  with the simplest cases of Gupta-Bleuler  theory of quantum  electromagnetic fields, as well as commonly used local renormalizable gauges (see, e.g.  \textsc{strocci}\cite[Chapter 6 and Appendix A.2]{Strocci})}.

\smallskip 
In accordance with Yukawa principle, a  positive mass gap is an experimental fact  for     weak and strong subnuclear forces:  \emph{A  limited force range  suggests its positive  mass} (cp. Yukawa 1949 Nobel  lecture). This   is a  \emph{quantum} effect  because, when the natural units of the  light velocity $c$ and  Planck quantum action  $\hbar$ are set to 1,  the energy-mass component of the relativistic energy-momentum vector has the natural physical dimension of the reciprocal  length $[L^{-1}]$. This is in spite of  the conformal symmetry of the Yang-Mills action functional on Minkowski spacetime (see, e.g., \textsc{glassey-strauss}\cite{Glassey}).

The infrared cutoffs provide the dimension $[L]$ for  the initially dimensionless Yang-Mills coupling constant (cp.  \textsc{faddeev}\cite{Faddeev-1} discussion of the reciprocal  effect  of ultraviolet cutoffs.) \footnote{Neither mechanism requires an auxiliary  Higgs field for providing a mass for Yang-Mills fields (despite of the 2012 LHC discovery of a quasi Higgs particle).}

\medskip
When  three  independent dimension units are chosen,  all physical magnitudes become  \emph{dimensionless}, a subject to mathematical calculus.

 The well known theorem that elliptic (pseudo)differenial operators on compact  manifolds have infinite discrete spectra (see, e.g. \textsc{shubin}\cite{Shubin})    may be a  hint for the quantum Yang-Mills energy-mass spectrum. However,
\begin{quotation}
Mathematically, quantum field theory involves integration, and 
elliptic operators, on infinite-dimensional spaces. Naive attempts 
to formulate such notions in infinite dimensions lead to all sorts of 
trouble. To get somewhere, one needs the very delicate constructions
considered in physics, constructions that at first sight look rather 
specialized to many mathematicians. For this reason, together with 
inherent analytical difficulties that the subject presents,
rigorous understanding has tended to lag behind development of physics.
 (\textsc{witten}\cite[p.346]{Witten})
\end{quotation}

This is in spite of the fact  that quantum field theory was born  immediately  after  quantum mechanics   that already matured to  a rigorous mathematical theory  in J. von Neumann's  monograph "Mathematische Grundlagen der Quantenmechanik" (1932). 

\smallskip
The concept   of  quantum field theory has double meaning itself:  
\begin{itemize}
\item Classical theory of quantum fields,  famously initiated by P. Dirac in 1927,
deals with \emph{operator-valued}  solutions of  \emph{non-linear} hyperbolic equations with  partial derivatives. Unfortunately, the non-bounded and non-commuting  linear operator values   seriously aggravate the  mathematical meaning  of non-lnear terms of the equations. Besides, the non-lnearity is incompatible with the quantum superposition Dirac postulate.
 
\item  Quantum theory of classical fields deals with  \emph{functional} solutions of  \emph{linear} 
Schr\"{o}dinger  equations with variational derivatives on  solution spaces of those hyperbolic equations.
There was a vivid discussion  among W. Heisenberg, P. Jordan,  and W. Pauli of the corresponding "Volterra mathematics". E.g.,   P. Jordan and W. Pauli  considered a 1-dimensional  variational Schr\"{o}dinger equation for  eigenfunctionals $\Psi(\phi(x))$ of massless scalar fields $\phi(x), x\in\mathbb{R}$  (\emph{Zur Quantumelectrodynamik ladungsfreier Felder,}  Zeitung  f\"{u}r Physik, \textbf{47} (1928))
\begin{equation*}
-\left(\frac{\hbar}{4\pi}\right)^2\int\!dx\, \left[
\frac{\delta^2}{\delta \phi(x)^2}
+ c^2\left(\frac{d\phi(x)}{d x}\right)^2\right]\Psi(\phi(x))=\lambda \Psi(\phi(x)).
\end{equation*}
\end{itemize} 
 By Bogoliubov-Shirkov-Schwinger  quantization postulate \cite[Chapter II]{Bogoliubov}, the  variational Schr\"{o}dinger operator is a quantization of the  dynamical energy-mass
invariant integral of the initial data for Euler-Lagrange equation for classical fields. \footnote{The proposed quantum Yang-Mills theory is of Lagrangian, not of Hamiltonian variety. It deals with Schwinger  quantization of Noether  energy-mass
functional, not with Faddeev-Popov  quantization of the constraint Hamiltonian dynamics.  Even so, by \textsc{moncrief}\cite{Moncrief},  there is an equivalence between   constrained Yang-Mills energy-mass functional   and constrained  Yang-Mills Hamiltonian  (cp.\textsc{faddeev-slavnov} \cite[Section 2, Chapter III]{Faddeev}) by itself.}

In 1954  \textsc{gelfand-minlos}\cite{Gelfand} proposed to solve variational quantum fields equations with  variational derivatives via approximations by solutions of partial differential equations with large but finite number of independent variables (cp. \textsc{berezin}\cite[Preface]{Berezin}). However the convergence of such approximations so far has  not been  established until the present paper.

\medskip
 The main content of the paper  is split into  three sections:
 
 \smallskip
Section 2 provides a mathematically rigorous context for a non-perurbative 
quantum field theory  on nuclear Kree-Gelfand triples. It includes the following items.
\begin{itemize}
  \item  Infinite-dimensional   symbolic  calculus of   operators with variational derivatives  as a mathematically rigorous  infinite-dimensional extension of formal \textsc{agarwal-wolf}\cite{Agarwal} symbolic  calculus  of partial   differential  operators. 
  \item Convergence of  approximations  of  operators with variational derivatives  by finite-dimensional  pseudodifferential operators, along with the convergence of  their  symbolic calculi.  This is a justification  for Gelfand-Minlos  method for solution \cite{Gelfand} of equations with variational derivatives.
  \item New theory of infinite-dimensional elliptic operators including their spectral properties. The ellipticity means the operator domination of  a power of the  number operator.  By Theorem \ref{pr:gap} the spectrum of an elliptic  non-negative operator is a sequence of non-negative eigenvalues converging to infinity. In particular it has positive a gap at the bottom.
\end{itemize}
Section 3 is an intricate application of the infinite-dimensional ellipticity   to  a quantum Yang-Mills theory:         
\begin{itemize}
\item    In the temporal gauge the   Yang-Mills system of the second order partial differential equations for Yang-Mills connections on four  dimensional Minkowski space is  equivalent to the Schwinger   semi-linear first order  hyperbolic  with  constraint  initial data on the Euclidean space  $\mathbb{R}^3$.

 \item By Ladyzhnskaya principle (\cite{Goganov}), the finite speed propagation   of solutions of   Yang-Mills system implies   the reduction of general Cauchy data to   Cauchy data with compact supports. By \textsc{goganov-kapitanskii}\cite{Goganov}, the Schwinger-Yang-Mills evolution system with compactly supported Cauchy data has unique   smooth  global solutions  on Minkowski space subject to necessary constraints. The restriction to compactly supported 
Cauchy data is equivalent to infrared cutoffs of their frequencies. 
 \item  Theorem \ref{pr:orthogonal} converts the non-linear manifold of constrained  initial data on a 3-dimensional euclidian ball   into a Gelfand triple of topological vector spaces. 
\end{itemize} 

Section 4  presents  the anti-normal quantization of the Yang-Mills energy-mass
functional in Gelfand-Kree triple over Coulomb quasi gauge. The Main Theorem \ref{pr:orthogonal} affirms the ellipticity of the non-negative anti-normal quantum energy-mass  Yang-Mills operator and, therefore, discreteness of its  spectrum. Proposition \ref{pr:ss}  exhibits an asymptotic self-similarity of the energy-mass spectra on  the  running energy-mass scale.\footnote{Famous 1900 Planck Ansatz that the classical energy  of confined electro-magnetic fields is proportional to their frequencies is  a historical artifact because of the first appearance of the (non-normalized) Planck constant.  That was before advent of  both  relativity and quantum theories. E. g. it does not imply the existence of  photons, another famous Einstein 1905 Ansatz. It is also a mathematical paradox since the \emph{classical} EM energy is a quadratic functional. Certainly, the discreteness  of the confined classical  EM energy does not imply  massive confined photons.}

\section{Elliptic  operators with variational derivatives }
 
\subsection{Review of Kree-Gelfand triples}
Consider  a   Gelfand    triple  of densely imbedded complex topological  spaces with the complex  conjugation $*$ (see, e.g., \textsc{gelfand-vilenkin}\cite{Gelfand-1})
\begin{equation}
\label{eq:Gelfand}
\mathcal{H}^\infty\ \subset\ \mathcal{H}^0\ \subset\ 
\mathcal{H}^{-\infty},
\end{equation} 
where 
\begin{itemize}
  \item The space $\mathcal{H}^0$ is a Hilbert space with a Hermitian sesquilinear form 
$z^*w$: \footnote{The notation is  bracketless as, e.g.,  in \textsc{berezin}\cite{Berezin}.}
  \item The space $\mathcal{H}^\infty$ of elements $z^*$ is a nuclear countably Hilbert space;
  \item The space $\mathcal{H}^{-\infty}$ of elements $w$ is the anti-dual space of $\mathcal{H}^\infty$ with respect to the Hermitian form $z^*w$.
\end{itemize}
Kree-Gelfand  nuclear triple $\mathcal{K}$ 
 (\textsc{kree} \cite{Kree-1} and \cite{Kree-2})  is a sesqui-holomorphic second   quantization of the Gelfand triple $\mathcal{H}$, with the induced   conjugation,  \begin{equation}
\label{eq:Kree}
\mathcal{K}^\infty
\ \subset\ \mathcal{K}^0
\ \subset\ \mathcal{K}^{-\infty}
\end{equation}
where (cp., e.g., \cite{Dynin-02})
\begin{itemize}
 \item  The space $\mathcal{K}^{-\infty}$ is the countably Hilbert  space of all entire holomorphic functionals $\Psi(z^*)$ on $\mathcal{H}^{\infty}$ with the topology of compact convergence. \footnote{A functional is entire  on a locally convex complex vector space if it is continuous and entire on every complex line in that space (see, e.g., \textsc{colombeau}\cite{Colombeau}).}

  \item The space $\mathcal{K}^0$ is  the   Bargmann-Hilbert space  of square integrable entire holomorphic functionals  on $\mathcal{H}^{-\infty}$ with respect to   the  Gaussian probability measure (see, e.g., \textsc{gelfand-vilenkin}\cite{Gelfand} ).  Bargmann Hermitian form   
   \begin{equation}
\label{eq:form}
 \langle\ \Psi\ |\ \Phi\ \rangle\  :=
\ \int \!dz^*dze^{-z^*z}\: \Psi^*(z)\Phi(z^*),  
 \end{equation}
where the $*$-dual $\Psi^*(z)$ is the complex conjugate of $\Psi(z^*)$.

  \item The space $\mathcal{K}^\infty$ is the anti-dual of $\mathcal{K}^{-\infty}$.

   \item    \emph{Borel-Fourier transform} (see, e.g.,  \textsc{colombeau}\cite[Chapter 7, Abstract]{Colombeau})
\begin{equation}
   \label{eq:basis}
\tilde{\Psi}(\zeta)\  :=
   \langle\ \Psi\ |\ e^\zeta\ \rangle, \quad  
 \Psi\in \mathcal{K}^{-\infty},
 \quad\quad 
 \tilde{\Psi}(z^*)\  :=
   \langle\ \Psi\ |\ e^{z^*}\ \rangle, \quad   
 \Psi\in\mathcal{K}^{\infty},
\end{equation}  
(where $ e^\zeta(z^*)= e^{z^*\zeta}=e^{z^*}$)
is a topological isomorphism between $\mathcal{K}^{-\infty}$ and the  nuclear space of entire functionals  $\Psi(\zeta)$ of exponential type on $\mathcal{H}^{-\infty}$. \footnote{An entire functional    is of exponential type if it has an exponential growth with respect to any continuous seminorm on $\mathcal{H}^{-\infty}$.}

\item  The Borel-Fourier transform intertwines directional differentiation and multiplication
\begin{equation}
\label{eq:int}
(\partial_{w^*}\Psi)\tilde{}\ =\ (w^*\zeta)\tilde{\Psi},\quad (\partial_{w}\Psi^*)\tilde{}\ =\ (\zeta^*w)\tilde{\Psi}^*.
\end{equation} 
\end{itemize}
By Grothendieck kernel theory, the nuclearity of the  Kree-Gelfand triples
 implies that the locally convex vector spaces  $\big(...\rightarrow ...\big)$ of continuous linear operators   are topologically isomorphic to the complete sesqui-linear  tensor products (both spaces are endowed with the topology of compact uniform convergence).
\begin{eqnarray}
\label{eq:general} \big(\mathcal{K}^\infty\rightarrow\mathcal{K}^{-\infty}\big)
 & \simeq &  \mathcal{K}^{*-\infty}\widehat{\otimes}\: \mathcal{K}^{-\infty},
 \\
\label{eq:tame} \big(\mathcal{K}^\infty\rightarrow\mathcal{K}^\infty\big)
 & \simeq &  \mathcal{K}^{*-\infty}
\widehat{\otimes}\:\mathcal{K}^\infty, 
 \\
\label{}  
\big(\mathcal{K}^{-\infty} \rightarrow\mathcal{K}^\infty\big)
 & \simeq &  \mathcal{K}^{*\infty}
\widehat{\otimes}\: \mathcal{K}^\infty,
\end{eqnarray}
where the  operators are \emph{tame} in the case of (\ref{eq:tame}). 

An operator  is  \emph{polynomial}  if its kernel  is a continuous polynomial on $\mathcal{K}^{*\infty}\times\mathcal{K}^\infty$,
so that polynomial operators are tame.
 
The formulas present the one-to-one correspondence  between operators and the  sesqui-holomorphic kernels of their matrix elements. 

The nuclear Gelfand  triple of the sesqui-Hermitian direct products   
\begin{equation}
\label{ }
\mathcal{H}^{*\infty}\times \mathcal{H}^{\infty}\ \subset\ 
\mathcal{H}^{*0}\times \mathcal{H}^{0}\ \subset\
\mathcal{H}^{*-\infty}\times \mathcal{H}^{-\infty}
\end{equation}
carries the Hermitian conjugation
\begin{equation}
\label{ }
(z^*,w)^*  :=
\ (w^*,z)
\end{equation} 
The associated Kree-Gelfand triples of  sesqui-holomorphic kernels consists of
\begin{equation}
\label{eq:kernels}
 \mathcal{K}^{*\infty}\widehat{\otimes} \mathcal{K}^\infty\ \subset   \ 
 \mathcal{K}^{*0}\widehat{\otimes}\mathcal{K}^0\ \subset
  \mathcal{K}^{*-\infty}\widehat{\otimes} \mathcal{K}^{-\infty}
\end{equation}
where  $\mathcal{K}^{*0}\widehat{\otimes}\:\mathcal{K}^0$ is the Hilbert space 
 of Hilbert-Shmidt kernels.

The corresponding exponential functionals are 
\begin{equation}
\label{eq:biex}
e^{(\zeta^*,\eta)}\big((z^*,w)^*\big)\ =\ e^{w^*\eta+\zeta^*z}.
\end{equation}

\subsection{Operators with variational derivatives }
Kree-Gelfand triple (\ref{eq:Kree}) has the  canonical linear  representation   by continuous linear transformations of $\zeta\in \mathcal{K}^{-\infty}
$ and $\zeta^*\in\mathcal{K}^{\infty}$ into  the adjoint operators of  \emph{creation and annihilation} continuous operators of multiplication and directional differentiation
\begin{eqnarray}
\hat{\zeta}:\ \mathcal{K}^\infty\ \rightarrow \mathcal{K}^\infty, & &
\hat{\zeta}\Psi(z^*)  :=
  (z^*\zeta)\Psi(z^*),\\
 \widehat{\zeta^*}:\ \mathcal{K}^{-\infty}\ \rightarrow \mathcal{K}^{-\infty}, & &
 \widehat{\zeta^*}\Psi(z)  :=
 (\zeta^*z)\Psi(z), \\
\widehat{\zeta^*}^\dag:\mathcal{K}^\infty\ \rightarrow\mathcal{K}^\infty,  & &\widehat{\zeta^*}^\dag\Psi(\zeta^*)   :=
  \partial_{\zeta^*}\Psi(z^*),\\
\hat{\zeta}^\dag:\ \mathcal{K}^{-\infty}
\ \rightarrow \mathcal{K}^{-\infty}, & &
 \hat{\zeta}^\dag\Psi(z)  :=
  \partial_\zeta\Psi(z),
\end{eqnarray} 
such that
\begin{enumerate}
   \item 
 Bosonic commutation  relation
 \begin{equation}
\label{eq:CCR}
[\widehat{\zeta^*}^\dag,\hat{\eta}]\ =\ \zeta^*\eta, \quad [\hat{\zeta}^\dag,\widehat{\eta^*}]\ =\ \eta^*\zeta.
\end{equation}

\item The exponentials $e^{\eta^*},\ \eta^*\in \mathcal{H}^{-\infty},$  and 
$e^{\eta},\ \eta\in \mathcal{H}^{\infty},$ are the eigenstates of the annihilation operators
\begin{equation}
\label{ }
\widehat{\zeta^*}^\dag e^\eta\ =\ (\zeta^*\eta)e^v,\quad \hat{\zeta}^\dag e^{\eta^*}\ =\ (\eta^*\zeta)e^{\eta^*}.
\end{equation}  
\end{enumerate}

Creators and annihilators generate  strongly continuous abelian operator groups of \emph{quantum exponentials} in 
$\mathcal{K}^{-\infty}
$ and $\mathcal{K}^{
-\infty}
$ parametrized by $\zeta$ and $\zeta^*$:
\begin{eqnarray}
\label{eq:1}
e^{\hat{\zeta}}:\mathcal{K}^{\infty}
\rightarrow\mathcal{K}^{\infty}
,\quad & &
e^{\hat{\zeta}}\Psi(z^*)\ =\ e^{z^*\zeta}\ \Psi(z^*);\\
\label{eq:2}
e^{\hat{\zeta}^\dagger}:\mathcal{K}^\infty\rightarrow\mathcal{K}^\infty,\quad & &
e^{\hat{\zeta}^\dagger}\Psi(z)\ =\Psi(z+\zeta);\\
\label{eq:3}
e^{\widehat{\zeta^*}}:\mathcal{K}^\infty\rightarrow\mathcal{K}^\infty,\quad & &
e^{\widehat{\zeta^*}}\Psi(z)\ =\ e^{\zeta^*z}\ \Psi(z);\\
\label{eq:4}
e^{\widehat{\zeta^*}^\dagger}:\mathcal{K}^{\infty}
\rightarrow\mathcal{K}^{\infty}
,\quad & &
e^{\widehat{\zeta^*}^\dagger}\Psi(z^*)\ =\  \Psi(z^*+\zeta^*).
\end{eqnarray}
 Borel-Fourier transform  of  kernels $M(\theta^*,\eta)$ 
\begin{equation}
\label{ }
\tilde{M}(z^*,w)\ =\ \langle\ M(\theta^*,\eta)\ | e^{z^*\eta+ \theta^*w}\ \rangle\ =\ 
\big\langle \ M(\theta^*,\eta)\ \big |\ \langle e^{z^*}|e^w\rangle\ \big\rangle
\end{equation}
 may be   quantized as   
  \emph{normal, Weyl,  and anti-normal pseudovariational  operators} $\widehat{M}$
  \begin{equation}
\label{ }
\widehat{M}:\ \mathcal{K}^{-\infty}
\rightarrow \mathcal{K}^{
-\infty},\quad
M\in\mathcal{K}^{*\infty}\hat{\otimes}\mathcal{K}^{\infty}\big)
\end{equation} 
  defined by their exponential matrix elements 
\begin{eqnarray}
\label{eq:n}
\langle\ e^{z^*}\ |\ \widehat{M}_\nu
\ |\  e^{w}\ \rangle\ &  :=
 & \big\langle \ M_\nu
(\theta^*,\eta)\  \big|\ \langle e^{z^*} |\ e^{\hat{\theta}}e^{\widehat{\eta^*}^\dagger}\ | e^{w}\ \rangle\ \big\rangle,
\\
\label{eq:w}
\langle\ e^{z^*}\ |\ \widehat{M}_\omega
\ |\ e^{w}\ \rangle\ &  :=
 & \big\langle\ M_\omega
(\theta^*,\eta)\  \big|\ 
\langle e^{z^*}| e^{\hat{\theta}+\widehat{\eta^*}^\dagger}| e^{w}\ \rangle\ \big\rangle,
\\
\label{eq:an}
\langle\ e^{z^*}\ |\ \widehat{M}_\alpha
\ |\ e^{w}\ \rangle\ &  :=
 & \big\langle\ M_\alpha
(\theta^*,\eta)\ \big|\ 
\langle e^{z^*}| e^{\widehat{\eta^*}^\dagger} e^{\hat{\theta}}| e^{w}\ \rangle\ \big\rangle.
\end{eqnarray}

\begin{proposition}
\label{pr:norm}
 Any continuous  linear operator $Q$ from $\mathcal{K}^{-\infty}$ to $\mathcal{K}^{-\infty}$ has a unique normal  co-kernel $M_\nu
^Q(\zeta^*,\eta)$.
\end{proposition}
\textsc{proof}\  
Since
\begin{eqnarray}
& & 
\langle\ e^{z^*}\ | e^{\hat{\eta}}e^{\widehat{\theta^*}^\dagger}\ | e^{w}\ \rangle\  =\
\langle\  e^{\hat{\eta}^\dagger}e^{z^*}\ |\ \widehat{\theta^*}^\dagger e^{w}\ \rangle \\
 & &
=\ \langle\  e^{z^*\eta}e^{z^*}\ |\  e^{\theta^*z}e^{w}\ \rangle\ =\ e^{z^*\eta+\theta^*z}e^{z^*w},
\end{eqnarray}
one has  
\begin{equation}
\label{eq:normal}
\langle\ e^{z^*}\ |\ \widehat{M}_\nu
\ |\ e^{w}\ \rangle\ =\ \langle\ M_\nu
(\theta^*,\eta)\ |\  e^{z^*\eta+\theta^*z}\rangle e^{z^*w}\ =\ \tilde{M}_\nu
(z^*,w)e^{z^*w}
\end{equation}
where $\tilde{M}_\nu
(z^*,w)$ is the sesqui-linear Fourier transform of  $M_\nu
(\theta^*,\eta)$ (see (\ref{eq:biex})). Thus $\tilde{M}_\nu
(z^*,w)e^{z^*w}$ is the kernel of $\widehat{M}_\nu
$.

By (\ref{eq:kernels}), any  Grothendieck kernel 
has a  unique such representation.\qed

\medskip
By the Taylor expansion centered at the origin, the sesqui-entire functionals   are uniquely  defined  by   their restrictions   to the real diagonal $(z^*,w=z)\in\Re\big(\mathcal{H}^{*\infty}\times \mathcal{H}^{\infty}\big)$, so that  the \emph{normal symbol}
of the operator $Q$
\begin{equation}
\label{eq:nsymb}
\sigma^Q_\nu
(z^*,z)\  :=
 \tilde{M}_\nu\ =\ \langle\ e^{z^*}\ |\ \widehat{M}_\nu\ e^z\ \rangle
^Q(z^*,z)
\end{equation}
exists and defines $Q$ uniquely.

\medskip
By Baker-Campbell-Hausdorff commutator formula and the canonical commutation relations (\ref{eq:CCR}),
 \begin{equation}
\label{eq:BCH}
e^{\hat{\eta}}e^{\widehat{\theta^*}^\dagger}\ =\ e^{\hat{\eta}+\widehat{\theta^*}^\dag}
e^{\theta^*\eta/2}, \quad  e^{\widehat{\theta^*}^\dagger}e^{\hat{\eta}}\ =\ 
e^{\hat{\eta}+\widehat{\theta^*}^\dag}e^{-\theta^*\eta/2}.
\end{equation}
Thus any operator $Q$ has Weyl and anti-normal co-kernels $M_\omega^Q$ and 
 $M_\alpha^Q$. Their restrictions to the real diagonal are  \emph{Weyl} and \emph{anti-normal}  symbols $\sigma_\omega^Q$  and $\sigma_\alpha^Q$ of $Q$.

By (\ref{eq:BCH}) the  symbols of the same operator $Q$ are related via \emph{Weierstrass transform}
(cp. \textsc{agarwal-wolf}\cite[formulas (5.29), (5.30), (5.31), page 2173]{Agarwal} in a finite dimensional case; \textsc{dynin} \cite{Dynin-02}) in white noise calculus):
\begin{eqnarray}
\label{eq:wn}
\sigma^Q_\omega
(z^*,z) & = &  e^{-(1/2) \partial_{z^*}\partial_z}\:\sigma^Q_\nu
(z^*,z),\\
\label{eq:ann}
\sigma^Q_\alpha
(z^*,z) & = & e^{- \partial_{z^*}\partial_z}\:\sigma^Q_\nu
(z^*,z), \\
\label{eq:wa}
\sigma^Q_\omega
(z^*,z) & = &  e^{(1/2) \partial_{z^*}\partial_z}\:\sigma^Q_\alpha
(z^*,z),
\end{eqnarray}
where the operator  $e^{\pm(1/2) \partial_{z^*}\partial_z}$  is the sesqui-linear Fourier equivalent  of the  multiplication by $e^{\pm(1/2)z^*z}$, i. e, 
\begin{equation}
\label{ }
 \partial_{z^*}\partial_z\  :=
\ \mbox{Trace}(\partial_{z^*}\partial_w).
\end{equation}

(Note that the multiplication by  $e^{z^*w}$ is a continuous operator in 
$\mathcal{K}^{-\infty^*}\hat{\otimes} \mathcal{K}^{-\infty}$, so that  the restrictions of  $e^{\partial_\zeta^*\partial_\zeta}\sigma(\zeta^*,\zeta)$ to the real diagonal  $\Re\big(\mathcal{H}^{*\infty}\times \mathcal{H}^{\infty}\big)$ are well defined.)

The following Proposition shows that anti-normal operators are infinite-dimensional
Berezin-Toeplitz integral operators from  $\mathcal{K}^\infty$ to 
$\mathcal{K}^{-\infty^*}$ (cp. \textsc{berezin}\cite[Equation (2.5)]{Berezin-72}).
\begin{proposition}
\label{pr:projection}
The  kernel of an anti-normal operator $Q=\widehat{M}_\alpha$  
\begin{equation}
\langle\ e^{z^*}\ |\ Q\ |\ e^{w}\ \rangle\ =\  e^{z^*w} \sigma_\alpha ^Q(w^*,w),
\end{equation}
i.e. $Q$ acts on $\Psi(w^*)$ as   the multiplication  $\sigma_\alpha
^Q(w^*,w)
\Psi(w^*)\in\mbox{Ent}\big(\mathcal{H}^{\infty})^*\times \mathcal{H}^{\infty}\big)$ followed by the orthogonal projection with the kernel $(e^{z^*w})^*=e^{w^*z}$ onto 
$\mbox{Ent}\big(\mathcal{H}^{\infty})$.
\end{proposition}
\textsc{proof}\ 
Since
\begin{eqnarray}
& & 
\langle\ e^{z^*}\ | e^{\widehat{\theta^*}^\dagger}e^{\hat{\eta}}\ | e^{w}\ \rangle\  =\
\langle\  e^{\widehat{\theta^*}}e^{z^*}\ |\ e^{\hat{\eta}} e^{w}\ \rangle \\
 & &
=\ \langle\  e^{\theta^*z}e^{z^*}\ |\  e^{z^*\eta}e^{w}\ \rangle\ =\ e^{z^*w}e^{\theta^*w+w^*\eta},
\end{eqnarray}
the  kernel (\ref{eq:an})
\begin{equation}
\langle\ M_\alpha
(\theta^*,\eta)\ |\ 
 e^{z^*w}e^{\theta^*w+w^*\eta}\ \rangle\ =\ e^{z^*w}\sigma_\alpha
(w,w^*).
\end{equation}
\qed
\begin{corollary}
\label{pr:Berezin}
The  diagonal matrix elements  of an anti-normal operator $Q$ on $\mathcal{K}^{\infty}$
\begin{equation}
\label{eq:Berezin}
\langle\ e^{z^*}\ |\ Q\ |\ e^{z}\ \rangle\ \geq\ \inf\:\sigma_\alpha ^Q(z^*,z),\quad
(z^*,z)\in \mathcal{H}^{\infty}.  
\end{equation}
\end{corollary}
This   Corollary is an infinite-dimensional extension of  Theorem 7.1 in \textsc{berezin}\cite{Berezin-72}.

\subsection{Quantized Galerkin approximations}
A \emph{Galerkin sequence}  $p_n,\ j=1,2,...,$ is an increasing  sequence  of tame orthogonal projections of rank $n$ strongly convergent to the identity operator in 
$\mathcal{H}^\infty$. The projectors $p_n$ are uniquely extended to the Galerkin families in $\mathcal{H}^0$ and $\mathcal{H}^{-\infty}$.  The notation  $p_n$ is kept for all  these extensions. 

The finite dimensional projectors induce the \emph{quantized Galerkin sequence} 
\begin{equation}
\label{ }
P_n\Psi(z^*)\  :=
\ \Psi(p_nz^*), \quad P_n\Psi(z)\  :=
\ \Psi(p_n z)
\end{equation}
of \emph{infinite  dimensional}  projectors in the triple $\mathcal{K}$ onto cylindrical triples isomorphic to the pulled back  sesqui-entire triples over the tautological finite dimensional  triple  
$\mathbb{C}^n\subset \mathbb{C}^n \subset \mathbb{C}^n$.

By Proposition \ref{pr:norm}, the   compressions of operators $Q_n :=
 P_nQP_n$  of $Q$ are  cylindrical pseudodifferential operators with the exponential kernels,
\begin{equation}
\label{ }
\langle\ e^{z^*}\ |\ Q_n\ | \ e^w\ \rangle\ =\ \langle\ e^{p_nz^*}\ |\ Q\ | \ e^{p_nw}\ \rangle,\end{equation}
i.e. pullbacks from $\mathbb{C}^j$ of finite dimensional   pseudodifferential operators of \textsc{agarwal-wolf}\cite{Agarwal}.

\begin{theorem}
\label{pr:cylindrical}
Operator $Q$ is the strong limit of the cylindrical pseudodifferential operators $Q_n$ on  $\mathcal{K}^\infty$.
\end{theorem}
\textsc{proof}\ 
The matrix element  $\langle\Psi^{*}|Q|\Phi\rangle$ is a separately continuous sesquilinear form on  the Frechet space $\mathcal{K}^\infty$. By a Banach theorem (see, e.g., \cite[v.1,Theorem V.7]{Reed}), the sesquilinear form is actually continuous 
on $\mathcal{K}^{*\infty}$. In particular,  operator $Q$ is the weak limit of $Q_n$ in $\mathcal{K}^{*\infty}$. Since   $\mathcal{K}^{*\infty}$ is a nuclear space,  the  weak convergence implies  the strong one in  the topology of $\mathcal{K}^{*\infty}$. 
 \qed

As $n\rightarrow \infty$, the exponential matrix elements
 \begin{equation}
\label{eq:convergence}
\langle\ e^{z^*}\ |\ Q_n\ | \ e^w\ \rangle\ =\ 
\langle\ e^{p_nz^*}\ |\  Q\ | \ e^{p_nw}\ \rangle\ 
\longrightarrow\ \langle\ e^{z^*}\ |\  Q\ | \ e^w\ \rangle,
\end{equation}
so that symbols of the cylindrical $Q_\nu$  converge  to the corresponding symbols of $Q$.

Thus if  operators  $Q_1$ and $Q_2$ are  tame then (cp. \textsc{agarwal-wolf}\cite[Theorem III.5]{Agarwal} 
\begin{corollary}
The symbols of the tame  $Q_3=Q_2Q_1$ are   convergent series (polynomials when $Q_1$ or $Q_2$ is a polynomial operator)
\begin{eqnarray}
& &
\label{eq:norm}
\sigma_\nu^{Q_3}(z^*,z)\ =\  \sum_{m=0}^\infty (m!)^{-1}\partial _z^m \sigma_\nu^{Q_2}(z^*,z)
\partial _{z^*}^m \sigma_\nu^{Q_1}(z^*,z),\\
 & &
 \label{eq:ant}
 \sigma_\alpha^{Q_3}(z^*,z) \ =\  \sum_{m=0}^\infty(m!)^{-1}
\partial _{z^*}^m \sigma_\alpha^{Q_2}(z^*,z)\partial _{z}^m 
\sigma_\alpha^{Q_1}(z^*,z),\\
& &
\label{eq:Wey}
\sigma_\omega^{Q_3}(z^*,z)\ =\  \sum_{m=0}^\infty (m!)^{-1}\Omega^m\big(\sigma_\omega^{Q_2}(z_2^*,z_2)
\sigma_\omega^{Q_1}(z_1^*,z_1)\big),
\end{eqnarray}
where $\Omega :=
\ (1/2)\big(\partial_{z_2^*} \partial_{z_1}\ -\  \partial_{z_1^*}\partial_{z_2}\big),\  z_2^* = z_1^*=z^*$.
\end{corollary}

Another consequence from Theorem \ref{pr:cylindrical} for essentially selfadjoint  
\begin{corollary}
\label{pr:spectrum}
Let $Q$ be the Friedrichs extension of a non-negative tame operator in 
$\mathcal{K}^{*\infty}$ to an (unbounded) selfadjoint operator in $\mathcal{K}^0$. Then the spectrum of $Q$ consists of limits of some spectral values of $Q_n$ as  $n$ converges to infinity. 
\end{corollary}
\textsc{proof}\  By Theorem  \cite[v.1, Theorem VIII.25(a)]{Reed} and Theorem \ref{pr:cylindrical},
the resolvent operators of $Q_n$ strongly converge to the   resolvent operator of $Q$.

Then  \cite[v.1,Theorem VIII.24(a)]{Reed} implies the lower semicontinuity of   the spectra    
convergence. \qed

\subsection{Elllipticity}
The tame quadratic \emph{number operator}  $N$ is defined by its quadratic symbols  
\begin{equation}
\label{eq:number}
\sigma^N_\nu\ =\ z^*z+1,\    \sigma^N_\alpha=z^*z,\ \sigma^N_\omega= z^*z+1/2.                                                 
\end{equation}
The notation  $N$ is transferred to the Friedrichs extension  of the number operator.
The eigenspaces $\mathcal{N}_n,\   n=\ 1,2,...,$ of $N$  with the  eigenvalues $n$ are the  spaces spanned by continuous  homogeneous polynomials of degree $n$. The number operator is positive and   self-adjoint in the Hilbert space $\mathcal{K}^0$. 

Thus $N$ has the simple  eigenvalue $\lambda_1=1$, and all other eigenvalues $\lambda_n=n$ form its essential spectrum.

A non-negative symmetric tame  operator $Q$  is  \emph{elliptic}
\footnote{The definitions of ellipticity in \cite{Bleher}, \cite{Lascar}, and 
\cite{Dynin-02} are not sufficient for the present paper.}
 if there exist  positive constants $r$ and  $c$ such that 
 \begin{equation}
\label{eq:domination}
\langle \Psi\ |\ Q\ |\  \Psi\ \rangle\ \geq\  c \langle\ \Psi^*\ |\ N^r\ |\ \Psi\ \rangle, \quad \Psi\in\mathcal{K}^\infty.
\end{equation} 
The notation  $Q$ is transferred to the Friedrichs extension  of the operator
$Q$.
\begin{theorem} \label{pr:gap}
The  spectrum of a positive elliptic operator $Q$ is a  sequence 
$\lambda_n(Q)\rightarrow +\infty$. In particular, Q  has a mass gap at the bottom of its spectrum.
\end{theorem}
\textsc{proof}\ 
Quantized Galerkin cutoffs $P_jQP_j$ of elliptic non-negative operators $Q$ in 
$\mathcal{K}^\infty$ are pullbacks of the  finite dimensional elliptic pseudodifferential operators  $Q_j$ on  $p_jX$.  The latter are non-negative and elliptic, so that they have discrete  spectra of eigenvalues $\lambda_n(Q_j)\geq 0$ converging to infinity (see, e.g. \textsc{shubin}\cite[Theorem 26.3]{Shubin}). Then   the operators $P_jQP_j$ have the  infinitely degenerated  spectra of $Q_J$. 

Similarly, for $k\geq j$, the   operators $p_jQ_kp_j$ on $p_kX$ are pullbacks of $Q_j$, so that $Q_j$  have trivial extensions $\breve{Q}_j$ of $Q_j$ (with $\breve{Q}_j=0$ on the orthogonal complement $(p_jX)^\perp$ in $p_kX$. Again, both $\breve{Q}_j$ and $Q_j$ have 
the same  spectra (though  with different but finite multiplicities).

Since $Q_k\geq \breve{Q}_j$,  a variational principle (see, e. g. \textsc{berezin-shubin}\cite[Appendix 1, Corollary 2 of Proposition 3.1]{Berezin-91})
\footnote{ If non-negative operators $A'\leq A''$ are essentially selfadjoint  on a common domain  $D$ with non-decreasing  sequential  discrete spectra $\lambda_n'$ and  $\lambda_n''$, then  $\lambda_n'\leq \lambda_n''$.}  
implies that the eigenvalues $\lambda_n(Q_j)$ monotonically increase with $j$. Then, by Corollary \ref{pr:spectrum}, the 
spectrum of the elliptic operator  $Q$ consists of some of the limits of these  monotonic sequences.

The cutoffs $P_jNP_j$ of the  number operator  are pullbacks of the finite dimensional pseudodifferential number  operators $N_j=p_jNp_j$ on $p_jX$. All they have the same spectra $\mathbb{Z}_+$, so that  $p_jcN^rp_j$ with positive $c$ and $r$  have discrete spectra  $cn^r$. Now  the same   variational principle and ellipticity condition (\ref{eq:domination}) imply that the eigenvalues $\lambda_n(Q_j)\geq cn^r$. Therefore, for any given integer $n$ there may be only finitely many limits of monotonic $\lambda_n(Q_j)$ as $j$ increases to infinity. \qed

\section{Quantum Yang-Mills theory}
\subsection{Classical Yang-Mills fields}
The \emph{global gauge  group}   $\mathbb{G}$ of a  Yang-Mills theory is   a  
connected semi-simple compact Lie group with the  Lie algebra $\mbox{Ad}(\mathbb{G})$. 

The notation $\mbox{Ad}(\mathbb{G})$ indicates that the Lie algebra carries the adjoint representation $\mbox{Ad}(g)X=gXg^{-1}, g\in\mathbb{G}, a\in Ad(\mathbb{G})$, of the group $\mathbb{G}$ and the corresponding self-representation $\mbox{ad}(X)Y=[X,Y],\ X,Y\in\mbox{Ad}(\mathbb{G})$.  Then $\mbox{Ad}(\mathbb{G})$ is identified with a Lie algebra of skew-
symmetric matrices and the matrix commutator as Lie bracket with  the \emph{positive  definite}  Ad-invariant  scalar  product
\begin{equation}
\label{eq:scalar}
X\cdot Y \  :=
\  \mbox{Trace}(X^TY), 
\end{equation}
where $X^T=-X$ denotes the matrix transposition. 

\bigskip
Let the Minkowski space $\mathbb{M}$ be oriented and time oriented with  the Minkowski metric signature $(-1,1,1,1)$. In a Minkowski coordinate systems 
$x^\mu, \mu=0,1,2,3$,  the metric tensor is diagonal.
  In  the natural unit system, the time coordinate $x^0=t$. Thus  $(x^\mu)=(t,x^i),\  i=1,\ 2,\ 3$. 
  
 The \emph{local gauge  group} $\mathcal{G}$ is the group of  infinitely differentiable $\mathbb{G}$-valued functions
 $g(x)$ on $\mathbb{M}$ with the pointwise group multiplication.  The  \emph{local gauge Lie algebra}  $\mbox{Ad}(\mathcal{G})$ consists of  infinitely differentiable $\mbox{Ad}(\mathbb{G})$-valued functions   on $\mathbb{M}$ with the pointwise Lie bracket.   
 
$\mathcal{G}$ acts via the pointwise adjoint action on $\mbox{Ad}(\mathcal{G})$   and correspondingly on  $\mathcal{A}$, the real vector space of \emph{gauge   fields}   $A=A_\mu(x)\in\mbox{Ad}(\mathcal{G})$. 

 \smallskip
   Gauge fields $A$ define   the \emph{covariant partial derivatives}  
   \begin{equation}\label{}
  \partial_{A\mu}X\  :=
\  \partial_\mu X- \mbox{ad}( A_\mu)X,\quad
X\in\mbox{Ad}(\mathcal{G}).  
\end{equation}
This definition shows that  in the natural units \emph{gauge  connections have the mass dimension} $1/[L]$. 

Any $g\in\mathcal{G}$ defines the affine \emph{gauge transformation}  
\begin{equation}\label{}
A_\mu\mapsto A_\mu^{g}:\ =\ \mbox{Ad}(g)A_\mu-(\partial_\mu g)g^{-1},\ A\in \mathcal{A},
\end{equation}
so that $A^{g_1}A^{g_2}=A^{g_1g_2}$.

\medskip
Yang-Mills \emph{curvature tensor} $F(A)$ is  the 
antisymmetric tensor \footnote{The dimensionless  Yang-Mills coupling  $\mathbf{g}^2_{{\tiny\mbox{YM}}}$ is set to 1}
\begin{equation}\label{}
F(A)_{\mu\nu} :=
\partial_\mu A_\nu-\partial_\nu A_\mu-[A_\mu,A_\nu].
\end{equation} 
The curvature is gauge invariant:
 \begin{equation}\label{}
\mbox{Ad}(g)F(A)\ =\ F(A^{g}), 
 \end{equation}
 as well as \emph{Yang-Mills Lagrangian} 
 \begin{equation}\label{}
 \label{eq:Lag}
  (1/4)F(A)^{\mu\nu}\cdot F(A)_{\mu\nu}.
  \end{equation}
 The corresponding gauge invariant  Euler-Lagrange equation is a   2nd order non-linear  partial differential equation $\partial_{A\mu}F(A)^{\mu\nu} =0$, called 
 the \emph{Yang-Mills equation} 
\begin{equation}
\label{eq:YM}
 \partial_\mu F^{\mu\nu}\ -\ [A_\mu, F^{\mu\nu}]\ =\ 0.
\end{equation}
 \emph{Yang-Mills fields} are solutions of Yang-Mills equation.

\subsection{First order formalism}
 In the temporal gauge  $A_0(t,x^k)=0$  \footnote{I. Segal theory (\cite{Segal-79}) of infinite-dimensional Sobolev Lie groups implies that for any infinitely differentiable gauge field on $\mathbb{M}$ there is a unique  infinitely  differentiable gauge transformation to  the temporal gauge.}  the   2nd order Yang-Mills equation (\ref{eq:YM})  is equivalent to  the 1st order Schwinger  hyperbolic system   for the time-dependent $A_j(t,x^k)$,  $E_j(t,x^k)$ on  $\mathbb{B}$ (see, e.g.,  \textsc{goganov-kapitanskii} \cite[Equation (1.3)]{Goganov})
\begin{equation}
\label{eq:evolution}
\partial_t A_k\  = \  E_k ,  \quad
\partial_tE_k \  = \  \partial_jF^j_k - [A_j,F^j_k],\ \quad\ F^j_k\ =\ \partial^j A_k - \partial_k A^j - [A^j,A_k].
\end{equation}
and the \emph{constraint  equations}
\begin{equation}
\label{eq:constraint}
 [A^k,E_k]    \ =\  \partial^kE_k, \quad \mbox{i.e.}\ \quad \partial_{k,A}E_k\ =\ 0.
\end{equation}
By  \textsc{goganov-kapitanskii} \cite{Goganov}, the evolution system is  a semilinear first order  partial differential  system  with  \emph{finite speed propagation} of the initial data, and the  Cauchy  problem for it with  constrained initial data at $t=0$ 
\begin{equation}
\label{ }
a_k(x)\  :=
\ A(0,x_k), \ e_k(x)\  :=
\ E(0,x_k), \quad \partial^ke_k=[a_k,e,_k]
\end{equation}
is \emph{globally and uniquely solvable} in local Sobolev spaces on the whole Minkowski space  $\mathbb{M}$ (with no restrictions at the space infinity.)

This fundamental theorem has been  derived via  Ladyzhenskaya  1949  method (see  \cite{Goganov}) by a reduction to the case of Cauchy data  on 3-dimensional balls  $\mathbb{B}=\mathbb{B}(R):\   |x|\leq R$. 

If the  constraint equations are satisfied  at $t=0$, then, in view of the evolution system, they are satisfied   for  all $t$ automatically. Thus the  \emph{1st order evolution system along with the  constraint equations for Cauchy data is equivalent  to the 2nd order Yang-Mills system}. Moreover the constraint equations are invariant under  \emph{time independent} gauge transformations. As the bottom line, we have
\begin{proposition}
In the temporal gauge Yang-Mills fields $A$ are in one-one correspondence with their  gauge transversal Cauchy data $(a,e)$ satisfying   the  equation $\partial_ae=0$.  
\end{proposition}

Consider the chain of Sobolev-Hilbert spaces $\mathcal{A}^s, -\infty<s<\infty,$ of (generalized) connections $a(x)$ on a     $\mathbb{B}$ of radius $R$  with respect to the norms 
\begin{equation}
\label{eq:SH}
 |a|_s^2 :=
\int_{\mathbb{B}}\! d^3x\,\big(a\cdot(1-\triangle)^sa\big) < \infty. 
\end{equation}

They define the real Gelfand nuclear triple (cp., e.g., \cite{Gelfand}) 
\begin{equation}
\label{eq:Gelf}
\mathcal{A}:\mathcal{A}^\infty\  :=
 \bigcap\mathcal{A}^s\ \subset \mathcal{A}^0\  \subset\
\mathcal{A}^{-\infty}  :=
\bigcup\mathcal{A}^s,
\end{equation}
where $\mathcal{A}^\infty$ is  a nuclear countably Hilbert space with the dual $A^{-\infty}$. 

Similarly we define the chain of Sobolev-Hilbert spaces $\mathcal{S}^s, -\infty<s<\infty,$ of (generalized) Lorentz scalar fields $u(x)$  on  $\mathbb{B}$ with values in $\mbox{Ad}\:\mathbb{G}$ and the Hilbert norms $|u|_s$. Let 
\begin{equation}
\label{eq:GelfS}
\mathcal{S}:\mathcal{S}^\infty\  :=
 \bigcap\mathcal{S}^s\ \subset \mathcal{S}^0\  \subset\
\mathcal{S}^{-\infty}  :=
\bigcup\mathcal{S}^s
\end{equation}
be the corresponding Gelfand triple.
 
Let   $a\in\mathcal{A}^{s+3},\ s\geq 0$. Then, by Sobolev embedding theorem $a$  is  continuously $s+2$-differentiable  on  $\mathbb{B}$ and, therefore, the following  gauged  vector calculus  operators are continuous:
\begin{itemize}
  \item \emph{Gauged gradient} $\mbox{grad}^a \ :\  \mathcal{S}^{s+1}\rightarrow  \mathcal{A}^s$,
 \begin{equation}
\label{ } 
\mbox{grad}^a_{k}u \  :=\  \partial_k u - [a_k,u].
\end{equation}

 \item \emph{Gauged divergence} $\mbox{div}^a\ :\  \mathcal{A}^{s+1}\rightarrow  \mathcal{S}^s$,
 \begin{equation}
\label{eq:div}  
\mbox{div}^a\:b \  := \   \mbox{div}b - [a;b], \quad  [a;b]:= a_k b_k.
\end{equation}
 \item \emph{Gauged curl}  $ \mbox{curl}^a\  :\ \mathcal{A}^{s+1}\rightarrow  \mathcal{A}^s$,
 \begin{equation}
 \label{eq:curl}  
 \mbox{curl}^ab \  :=\ \mbox{curl}\:b - [a\stackrel{\times}{,}b], \quad  [a\stackrel{\times}{,}b]_i\  :=\ \varepsilon_{ijk}\ [a_j,b_k]. 
 \end{equation}
  \item \emph{Gauged Laplacian}  $\triangle^a:\ \mathcal{S}^{s+2}\rightarrow \mathcal{S}^s$,
  \begin{equation}
  \label{eq:Laplacian}  
\triangle^a u\  :=\ \mbox{div}^a(\mbox{grad}^au).
  \end{equation}
\end{itemize} 
 The  adjoints of the gauged   operators  are
\begin{equation}
\label{eq:adjoint}
 (\mbox{grad}^a)^*  =\ -\mbox{div}^a, \quad \mbox{curl}^{a*}=-\mbox{curl}^a.
\end{equation}
\begin{lemma}
\label{pr:sur}
If  $a\in\mathcal{A}^{s+3}, s\geq 0,$ then  the  operator  $\mbox{div}^a:\mathcal{A}^{s+1}\rightarrow \mathcal{A}^s$ is surjective.
\end{lemma}
\proof
Let $\mathring{\mathcal{S}}^{s+2},\ s\geq 0,$  denote  the closure in $\mathcal{S}^{s+2}$ of the space of $a$'s with compact support in the interior of $\mathbb{B}$.
The conventional Laplacian  $\triangle^0:\ \mathring{\mathcal{S}}^{s+2}\rightarrow \mathcal{S}^s$ is an isomorphism
(see e.g. \textsc{agranovich et al} \cite{Agranovich}).

The gauged Laplacian $\triangle^a$   differs from the usual  Laplacian $\triangle^0$ by first order differential operators, and, therefore is a Fredholm operator of zero  index from $\mathring{\mathcal{S}}^{s+2}$ to 
$\mathcal{S}^s,\ s\geq 0$.

If $\triangle^au=0$ then then  $(\triangle^au)^* u =  (\mbox{grad}^a\,u)^* (\mbox{grad}^a\,u)$ , so that  $\mbox{grad}\,u=[a,u]$.
The computation
\begin{equation}
\label{ }
(1/2)\partial_k(u\cdot u)= (\partial_ku\cdot u)=[a_k,u]\cdot u= -\mbox{Trace}(a_kuu-ua_ku)]=0
\end{equation}
shows that  the solutions    $u\in\mathring{\mathcal{S}}^{s+2}$ are constant. Because they vanish on the ball boundary, they vanish on the whole ball.   Since the index of the Fredholm operator $\triangle^a$ is zero,  its range  is a closed subspace with the codimension equal to the dimension  of its  null space. Thus    the operator $\mbox{div}^a\mbox{grad}^a$  is surjective , and so is $\mbox{div}^a$. \qed

\subsection{Coulomb quasi gauge}
 Consider the  bundles $\mathcal{C}^s, s\geq 0$ of constraint Cauchy data   with the base  
$\mathcal{A}^{\infty}$ and the null spaces $\mathcal{E}^{s+1}_a$ of the operators $\mbox{div}^a:\ \mathcal{E}^{s+1}\rightarrow \mathcal{E}^{s}$ as fibers over  $a\in\mathcal{A}^{\infty}$. 
  
   Their intersection $\mathcal{C}^\infty$ is  a bundle of nuclear countably Hilbert spaces over the nuclear countably Hilbert base   $\mathcal{A}^{\infty}$.  Together with the unions of the dual spaces $\mathcal{C}^{-s}$ they form  a bundle of nuclear Gelfand triples $\mathcal{C}$ over the same base.
\begin{theorem}
\label{pr:orthogonal}
The  bundle  $\mathcal{C}^\infty$ is  smoothly trivial,  so that the total space of  $\mathcal{C}^\infty$ is smoothly isomorphic to the direct product  of  its base  $\mathcal{A}^{\infty}$ and the  fiber 
$\mathcal{C}^{\infty}_{a=0}$, the nullspace of the operator $\mbox{div}$ in $\mathcal{E}^{\infty}$. 
\end{theorem}
\proof
For $0\leq s\leq \infty$ consider the mapping 
\begin{equation}
\label{ }
f:\ \mathcal{A}^{s+2}\times  \mathcal{E}^{s+1}\rightarrow \mathcal{A}^{s},\quad 
f(a,e) :=
 \mbox{div}_a(e)
\end{equation}
Sobolev imbedding theorem shows  that the mapping is continuous.
Lemma \ref{pr:sur} implies  that the continuous  partial Frechet derivatives $\partial_ef(a,e)$
are  bounded  linear operators onto a fixed Hilbert space $T(s)$, the orthogonal complement of constant $a$'s.
 continuously dependent on the parameter $a \in \mathcal{A}^{s+2}$.  The  restrictions of 
 $\partial_ef(a,e)$ to the  orthogonal complements of the null spaces of  $\mbox{div}_a$ are one-to-one. By  the  implicit function theorem on Hilbert spaces (see, e.g.,  \cite{Lang}), this implies that  the explicit solutions $e=e(a)$ of the equation  $f(a,e)=0$ provide  infinitely smooth local  trivializations  of  Hilbert bundles $\mathcal{C}^s$. 

Their  intersection $\mathcal{C}^\infty=\cap\mathcal{C}^s$ is a locally trivial $C^\infty$-bundle over 
$\mathcal{A}^\infty$ with the associated locally trivial bundle of smooth $*$-orthonormal frames in the fibers.

Since $\mathcal{A}^\infty$ is a Frechet space, its smooth homothety retraction to the origin $a=0$ has a homotopy lifting to the frame space. Thus the bundle $\mathcal{C}^\infty$ is trivial\footnote{Cp. \textsc{booss-bleecker}\cite[page 67]{Booss}} , so that the total set of constraint Cauchy data carries the global chart $\mathcal{A}^\infty \times \ \mathcal{C}^{\infty}_{a=0}$.\qed

\medskip
Let $\dot{\mathcal{A}}^{s}$ and  $\dot{\mathcal{E}}^{s}$ denote the nullspaces of the operator 
$\mbox{div}$ in $\mathcal{A}^s$ and $\mathcal{E}^s$.

By \textsc{dell'antoniio-zwanziger}\cite{Dell'Antonio}, the closures of  smooth gauge orbits in $\mathcal{H}^0 :=
\mathcal{A}^{0}$ intersect $\dot{\mathcal{A}}^{0}$. These closures  are the orbits of  the Sobolev group, the closure in  Sobolev space $W^{1,2}(\mathbb{B})$ of the group  of smooth gauge transformations.  (The Sobolev group  is  a topological group of continuous transformations in $\mathcal{A}^0$.)
Thus  $\mathcal{H}^0 :=
 \dot{\mathcal{A}}^0\times \dot{\mathcal{E}}^0$ is a \emph{quasi-gauge} for the orbifold of the direct product of the  parallel transports\ (i.e. every  $(a,e)\in\mathcal{H}^0$ is on an  orbit but some orbits may intersect $\mathcal{H}^0$ more than once (cp. \textsc{singer}\cite{Singer} and \textsc{narasimhan-ramadas}\cite{Narasimhan}). 
\medskip 
The Gelfand triple 
\begin{equation}
\label{eq:Coulomb}
 \mathcal{H}:\  \mathcal{H}^\infty  :=
 \dot{\mathcal{A}}^\infty\times \dot{\mathcal{E}}^\infty\ \subset\
 \mathcal{H}^0  :=
\dot{\mathcal{A}}^0\times \dot{\mathcal{E}}^0\ \subset\
\mathcal{H}^{-\infty}  :=
 \dot{\mathcal{A}}^{-\infty}\times \dot{\mathcal{E}}^{-\infty}
\end{equation}
 is the direct product of the Gelfand triples $\dot{\mathcal{A}}$ and  $\dot{\mathcal{E}}$.

\subsection{Quantum Yang-Mills energy-mass spectrum}
The  Noether  energy-mass functional of smooth Yang-Mills Cauchy data  (cp.\textsc{glassey-strauss}\cite[Section 3]{Glassey}) on  $\mathbb{B}$ is
\begin{equation}
\label{eq:Noether}
M(a,e)\  :=
\ (1/2)\int_{\mathbb{B}}\!d^3x\:
\big((\mbox{curl}\,a -  [a\stackrel{\times}{,}a])\cdot (\mbox{curl}\,a -  [a\stackrel{\times}{,}a])\ +\ e\cdot e \big).
\end{equation} 
The density
$(\mbox{curl}\,a -  [a\stackrel{\times}{,}a])\cdot (\mbox{curl}\,a -  [a\stackrel{\times}{,}a])$
 is  the scalar gauge curvature of $a$ and, as such, is invariant under the  gauge parallel transport but the density $e\cdot e$ is not.

 At the same time the density $e\cdot e$ is  invariant under the flat isometric parallel transport provided by Theorem \ref{pr:orthogonal}.
Thus the energy-mass
functional $M$ is constant on smooth  orbits of the direct product  of both parallel transports.

\medskip
Convert the Coulomb quasi-gauge  triple \ref{eq:Coulomb} into the  complex Gelfand triple $\mathcal{H}$  with conjugation where 
the real and imaginary parts are the direct factors
\begin{equation}
\label{ }
\Re\mathcal{H}\  :=\  \dot{\mathcal{A}},\quad \Im\mathcal{H}\  :=\  \dot{\mathcal{E}}.
\end{equation}
Let  the polynomial  energy-mass
functional $M(z,z^*) := M(a,e)$ (ref:eq:Noether) be  the \emph{anti-normal symbol} of 
the tame quantum Yang-Mills energy-mass
operator $\widehat{M}_\alpha:\mathcal{H}^\infty\rightarrow\mathcal{H}^\infty$. 
\begin{theorem} \label{pr:main}
The (unique) non-negative selfadjoint Friedrichs extension of  $\widehat{M}_\alpha$ in 
$\mathcal{M}^0$ is elliptic. Its spectrum is an  infinite   sequence of non-negative eigenvalues. In particular, the spectrum has  a positive mass gap.
\end{theorem}
\textsc{proof}\  
(A)\   
\begin{eqnarray*}
\label{eq:expansion}
& &
\int_{\mathbb{B}}\!d^3x\:
\big(\mbox{curl}\:a- [a\stackrel{\times}{,}a]\big) \cdot \big( \mbox{curl}\:a- [a\stackrel{\times}{,}a]\big)\\
& &
=\ \int_{\mathbb{B}}\!d^3x\:
\big(\mbox{curl}\:a\cdot \mbox{curl}\:a\ +\ ( [a\stackrel{\times}{,}a])\cdot( [a\stackrel{\times}{,}a])\\
& &
-\ 2 \int_{\mathbb{B}}\!d^3x\: \mbox{curl}\:a\cdot( [a\stackrel{\times}{,}a]).
\end{eqnarray*}
 The vector identity
  \begin{equation*}
\mbox{curl}( [a\stackrel{\times}{,}b])\ =\  (\mbox{div}\:a)b\ -\ a(\mbox{div}b)\ +\  (b\cdot \nabla)a\ -\  (a\cdot \nabla)b
  \end{equation*}
  (where $\nabla$ is the gradient vector) converts the last integral after integration by parts into
   \begin{equation}
 -\int_{\mathbb{B}}\!d^3x\: a\cdot\mbox{curl}( [a\stackrel{\times}{,}a])\ =\ 
  -\int_{\mathbb{B}}\!d^3x\: [\mbox{div}\: a,a].
 \end{equation}  
Therefore, by (\ref{eq:scalar}), the energy-mass functional (\ref{eq:Noether})
on $\mathcal{H}^\infty$ (so that $\mbox{div}\,a =0$) becomes
   \begin{equation}
\label{eq:equal}
M(a,e)\  =\  (1/2)\int_{\mathbb{B}}\!d^3x\:\big(\mbox{curl}\:a\cdot \mbox{curl}\:a\ +\  [a\stackrel{\times}{,}a]\cdot [a\stackrel{\times}{,}a]\ +\ e\cdot e\big).
\end{equation}  

\smallskip
\noindent  (B)\ Let  $b_i$ be a basis for $\mbox{Ad}\!(\mathbb{G})$ with $b_i\cdot b_j=\delta_{ij}$.
Then, since \emph{ the   gauge group $\mathbb{G})$ is a simple Lie group}, the structure constants $c^k_{ij}\ =\ [b_i,b_j]\cdot b_k$ are anti-symmetric under interchanges of  all $i,j,k$.  
Thus if   $a=\alpha^ib_i$ then (see \textsc{simon}\cite[page 217]{Simon})
\begin{eqnarray}
& &
( [a\stackrel{\times}{,}a])\cdot( [a\stackrel{\times}{,}a])\ =\  
\alpha^i\alpha^j\alpha^l\alpha^m\varepsilon^{ijk}([b_i,b_j]\cdot b_k) \varepsilon^{lmk}([b_l,b_m]\cdot b_k)\\
& &
 =\ \sum_k\alpha^i\alpha^j\alpha^l\alpha^m c^k_{ij}c^k_{lm}\ =\  \sum_k(\alpha^i\alpha^jc^k_{ij})^2.
\end{eqnarray}
Because $i\neq j$,  the Laplacian $\partial_a^2\ :=\
\sum_j\partial^2/\partial_{\alpha_j}\partial_{\alpha_j}$ applied to $[a\stackrel{\times}{,}a])\cdot( [a\stackrel{\times}{,}a]$ produces
\begin{equation}
\label{eq:Killing}
 2\sum_k(\alpha^i\ c^k_{ij})(\alpha^j c^k_{ij})\ = \ -2K(a,a)\ = c (a\cdot a)>0
\end{equation}
where $K(a,a)$ is the negative definite Killing quadratic form on $\mbox{Ad}\!(\mathbb{G})$ (see 
\textsc{simon}\cite[Equation (13)]{Simon}
and $c$ is a (positive) constant, by simplicity of  the gauge group $\mathbb{G}$.

\medskip
\noindent (C)\ Since   $\partial_z^*\partial_z =  \partial_a^2\ +\ \partial_{e}^2$,
\begin{equation}
\label{eq:Laplacian}
e^{\partial_z^*\partial_z/2}\big(( [a\stackrel{\times}{,}a])
\cdot ( [a\stackrel{\times}{,}a])\big)\ \stackrel{(\ref{eq:Killing})}{=}\ ( [a\stackrel{\times}{,}a])\cdot ( [a\stackrel{\times}{,}a])\ +\ \frac{c}{2}a\cdot a + c.
\end{equation}
Therefore   the Weyl symbol of the operator $H:=\widehat{M}_\alpha$
\begin{equation}
\label{ }
\sigma^H_\omega(a,e)\ \stackrel{(\ref{eq:wa}),(\ref{eq:Laplacian})}{=}\ 
\int_{\mathbb{B}}\!d^3x\:\big(( [a\stackrel{\times}{,}a])
\cdot ( [a\stackrel{\times}{,}a])\ +\ \ e\cdot e\ +\  \frac{c}{2}a\cdot a \ + c+\frac{1}{2}\big).
\end{equation}

 The Weyl quantization of  $( [a\stackrel{\times}{,}a])\cdot ( [a\stackrel{\times}{,}a])$  is the non-negative  operator of multiplication with $( [a\stackrel{\times}{,}a])\cdot ( [a\stackrel{\times}{,}a])\geq 0$  in the "$(a,e)$-representation" of the canonical commutation relations (cp. \textsc{agarwal-wolf} \cite[Section VII, page 2177]{Agarwal}). 

\medskip
\noindent (D)\ By (\ref{eq:number}),  
$$\int_{\mathbb{B}}\!d^3x\:(a\cdot a \ +\ e\cdot e\ +\ 1/2)
$$  
 is the anti-normal symbol of the number operator $N$.

Altogether we get the operator inequality of non-negative operators
\begin{equation}
\label{eq:B}
\widehat{M}_\alpha\ \geq\ C\widehat{N}_\alpha
\end{equation}
where $C$ is a positive constant.

Now the spectral Theorem \ref{pr:gap} implies the  Theorem \ref{pr:main}. \qed 

The restriction to the periodic  Yang-Mills fields is equivalent  to infrared cutoffs.
\begin{proposition}
\label{pr:ss}
The spectra of quantum Yang-Mills  energy-mass
operators  are  self-similar in the   inverse  proportion to  the radius of the ball $\mathbb{B}(R)$.
 \end{proposition}
\textsc{proof}\ 
The scaling transformation $(x,\ a,\ e)\ \mapsto (x/R,\ a/R,\   e/R^2)$
converts the energy-mass
functional (\ref{eq:Noether}) over  $\mathbb{B}(R)$  of the radius $R$
\begin{equation}
\label{ }
(1/2)\int_{\mathbb{B}(R)}\!d^3x\:
\big((\mbox{curl}\,a -  [a\stackrel{\times}{,}a])\cdot (\mbox{curl}\,a - [a\stackrel{\times}{,}a])\ +\ e\cdot e \big)
\end{equation}
 into the scaled energy-mass functional over the unit ball $\mathbb{B}(1)$
\begin{equation}
\label{ }
(1/2R)\int_{\mathbb{B}(1)}\!d^3x\:
\big((\mbox{curl}\,a -  [a\stackrel{\times}{,}a])\cdot (\mbox{curl}\,a -  [a\stackrel{\times}{,}a])\ +\ e\cdot e \big)
\end{equation}
The modified scalar product 
\begin{equation}
\label{ }
\int_{\mathbb{B}(1)}\!d^3x\:(a\cdot a/R\ +\ R e\cdot e)
\end{equation}
is invariant under the scaling, so that the quantum canonical relations are conserved under the scaling. \qed 
\begin{remark}
The  quantum Yang-Mills energy-mass excitations   above  the mass gap may indicate the existence of infinitely many massive glueballs. 
\end{remark}

\end{document}